\documentclass[12pt]{article}
\usepackage[a4paper, margin=0.5in]{geometry}
\usepackage{authblk}
\usepackage{amssymb}
\usepackage{latexsym}

\usepackage{amsmath,mathrsfs}
\usepackage{amsthm}
\usepackage{algorithm,algorithmic}
\usepackage{graphicx}
\newtheorem{theorem}{Theorem}[section]
\newtheorem{remark}{Remark}[section]
\usepackage{chngcntr}
\counterwithin{table}{section}
\counterwithin{figure}{section}

\usepackage{float}
\usepackage{lineno}
\usepackage{color}
\usepackage{enumerate}

\newcommand\modify[1]{{#1}}

\begin{document}

\title{A gradient flow model for ground state calculations in Wigner formalism based on density functional theory}
\date{}
\author[1]{Guanghui Hu}
\author[2,3]{Ruo Li}
\author[4]{Hongfei Zhan\thanks{Corresponding Author}}
\affil[1]{State Key Laboratory of Internet of Things for Smart City and Department of Mathematics, University of Macau, Macao SAR, 999078, People's Republic of China}
\affil[2]{CAPT, LMAM and School of Mathematical Sciences, Peking University, Beijing, 100871, People's Republic of China}
\affil[3]{Chongqing Research Institute of Big Data, Peking University, Chongqing, 401121, People's Republic of China}
\affil[4]{School of Mathematical Sciences, Peking University, Beijing, 100871, People's Republic of China}
\maketitle

\begin{abstract}
	In this paper, a gradient flow model is presented for conducting ground state calculations in Wigner formalism of many-body system in the framework of density functional theory. 
	Theoretically, an energy functional in the Wigner formalism is proposed, based on which the minimization problem is designed and analyzed for the ground state, providing a new perspective for ground state calculations of the Wigner function.
	Employing density functional theory, a gradient flow model is built upon the energy functional to obtain the ground state Wigner function representing the entire many-body system. 
	Numerically, a parallelizable algorithm is developed using the operator splitting method and the Fourier spectral collocation method, whose numerical complexity of single iteration is $O(n_{\rm DoF}\log n_{\rm DoF})$.
	Numerical experiments demonstrate the anticipated accuracy, encompassing the one-dimensional system with up to $2^{21}$ particles and the three-dimensional system with defect, showcasing the potential of our approach to large-scale simulations and computations of systems with defect.
	To the best of our knowledge, this is the first deterministic method for calculating the ground state Wigner function of the three-dimensional many-body system.
		
\end{abstract}

\section{Introduction}\label{sec:introduciton}

\modify{In the Schr\"odinger formalism, density} functional theory \cite{sham1966one} (DFT) is one of the most widely used approximate model in electronic structure calculations \cite{parr1995density,saad2010numerical,lin2019numerical}. Theoretically, it has been proved that three dimensional ground state electron density is a fundamental quantity in a given many-body system \cite{hohenberg1964inhomogeneous}. Numerically, the ground state density is often determined by either energy minimization or solving the self-consistent Kohn-Sham equations. Despite the diversity of numerical methods for solving the ground state in Schr\"odinger formalism, significant challenges arise in large-scale simulations, encompassing algorithm scalability \cite{motamarri2014subquadratic,nakata2020large,das2022dft}, convergence issue \cite{zhang2015many} and even the memory requirement for depicting the whole system. It is worth mentioning that Das et al achieved ab-initio simulations of quasicrystals and interacting extended defects in metallic alloys consisting of hundreds of thousands electrons in \cite{das2023large}. However, the sheer number of orbitals that need to be solved remains a fundamental problem for large-scale simulations. The Wigner formalism offers an alternative approach to both theoretically describe and numerically solve quantum systems, holding the potential for efficiency simulations of large-scale systems.

In contrast to the conventional Schr\"odinger wave function in Hilbert space, the Wigner phase-space quasi-distribution function \cite{wigner1932} provides an equivalent approach to describe quantum object that bears a close analogy to classical mechanics \cite{zachos2002}. The Wigner formalism has been applied to a variety of situations ranging from atomic physics \cite{vacchini2007} to quantum electronic transport \cite{schwaha2013,weinbub2018} and many-body quantum systems \cite{sellier2015introduction,xiong2016advective}. Furthermore, both pure and mixed states of a quantum system can be handled in a unified manner by Wigner function \cite{william2008}. Hence, in the context of density functional theory, a set of Kohn-Sham orbitals can be depicted by a single mixed state Wigner function, which sums up all the pure state Wigner functions corresponding to each orbital. This facilitates the ground state calculations based on a single Wigner function instead of multiple wave functions for Kohn-Sham orbitals. All these features motivate the research on developing models and numerical methods to determine the Wigner functions of a quantum system.

Different from the situation for Schr\"odinger equation that there have been a number of mature approximate models and numerical methods, more efforts are required towards the Wigner functions. 
The first attempts to simulate quantum phenomena by the Wigner function were \cite{frensley1987,frensley1990} for one-dimensional one-body systems. Subsequently, Wigner simulations were accomplished using the spectral collocation method combined with the operator splitting method \cite{arnold1994mixed,arnold1995operator}. Recently, several methods were designed for the simulation based on Wigner function, including the spectral element method \cite{shao2011,xiong2016advective,chen2019high,xiong2023characteristic}, the spectral decomposition \cite{van2017efficient}, the moment method \cite{cai2013,li2014hyperbolic,zhan2022wigner}, the discontinuous Galerkin method \cite{gamba2009adaptable}, the Gaussian beam method \cite{yin2013}, the weighted essentially non-oscillatory scheme \cite{dorda2015}, and the exponential integrator \cite{furtmaier2016semi}. Additionally, various stochastic methods have been developed, such as the signed particle Wigner Monte Carlo method \cite{nedjakov2004,nedjalkov2013,sellier2014benchmark} and path integral method \cite{larkin2016,larkin2017,amartya2019}. 
In many-body scenarios, the Wigner based simulations have been achieved by the advective-spectral-mixed method \cite{xiong2016advective}, the Monte Carlo method \cite{sellier2014many,sellier2015fermion,sellier2016full} and the method based on branching random walk \cite{shao2020}. 
For dynamic studies of a given system, an initial state should be specified, which is typically the ground state. Nemerous theoretical works have derived ground state Wigner functions for specific models, such as the hydrogen atom \cite{dahl1982}, the closed-shell atom \cite{springborg1987}, and the Moshinsky atom \cite{dahl2009}. In terms of numerical computations, Sellier and Dimov proposed a viable framework in \cite{sellier2014} for addressing both time-dependent and time-independent problems of three-dimensional systems using the Monte Carlo method. On the contrary, deterministic methods are limited to one-dimensional systems, encompassing the modified tau spectral method \cite{hug1998fundamentals,hug1998generalization} and the Fourier transform \cite{bondar2016}. In our previous work \cite{zhan2022wigner}, ground state calculations in Wigner formalism for DFT examples were accomplished using simplified Grad moment method combined with an imaginary time propagation method. However, deterministic approaches for ground state calculations in Wigner formalism with density functional theory remain limited to two-body systems.

In this paper, in the category of deterministic approach, a gradient flow model is proposed for ground state calculations in Wigner formalism of many-body systems within the framework of density functional theory. In particular, a gradient flow model for one-body systems is firstly derived. Introducing density functional theory, this model is extended to many-body systems by solving a single mixed state Wigner function. To enhance computational efficiency, an operator splitting scheme is adopted to decompose the original system into three sub-equations, enabling parallel implementation. Based on the structure of derived equations, the Fourier pseudo-spectral method is employed for discretization in both $\mathbf{x}$ and $\mathbf{p}$ directions. Consequently, each sub-equation can be solved with $O(n_{\rm DoF})$ computational cost for degree of freedom (DoF) number $n_{\rm DoF}$, resulting in an overall computational cost of single time iteration as $O(n_{\rm DoF}\log n_{\rm DoF})$. To validate the proposed method, two toy models are presented, which are generated by the periodic extension of two-body systems. The first example is a one-dimensional delta-interacting system with a local density approximation, where the ground state for the system with up to $2^{21}$ electrons are calculated. The second example is a three-dimensional system with Coulomb interaction, including a scenarios with a defect. Anticipated spectral accuracy can be successfully observed in all the computations. And these examples demonstrate the potential applications of our approach to large-scale systems and systems with defects.

\modify{
Our contributions of this work can be summarized as follows.
\begin{enumerate}[1.]
	\item The energy functional for the ground state in Wigner formalism is proposed, providing a new perspective for ground state calculations in Wigner formalism.
	\item A gradient flow model is presented for ground state calculations in Wigner formalism of many-body systems with the aid of density functional theory. To our best knowledge, our method is the first deterministic approach solving three-dimensional many-body ground states in Wigner formalism.
	\item Efficient algorithm is developed for the model, whose computational cost of single iteration is $O(n_{\rm DoF}\log n_{\rm DoF})$ for DoF number $n_{\rm DoF}$.
	\item Numerical simulations successfully deliver the desired results for the one-dimensional system with up to $2^{21}$ particles and the three-dimensional system with defect, fully demonstrating the descriptive capability of the Wigner formalism, and showcasing the potential of our approach to large-scale simulations and computations of systems with defect.
\end{enumerate}
}

The rest of this paper is organized as follows: In Sect. \ref{sec:wigner}, the eigenvalue problem in Wigner formalism is briefly introduced, based on which a gradient flow model for one-body systems is derived. Incorporating density functional theory, the gradient model is extended to many-body systems in Sect. \ref{sec:dft}. Sect. \ref{sec:numerical method} discussion the implementation of an operator splitting method and the Fourier pseudo-spectral method for efficient simulations of the model. Numerical results are demonstrated in Sect. \ref{sec:numerical result}. Finally, a conclusion of this paper and the discussion of future work are provided in Sect. \ref{sec:conclusion}.

For simplicity, the Hartree atomic unit is adopted hereafter, i.e., take $\hbar=m=e=1$ for reduced plank constant, electron mass and electron charge.

\section{Wigner function of ground state}\label{sec:wigner}
In this section, the Wigner formalism and the Wigner eigenvalue problem are introduced. Subsequently, an energy function in the Wigner formalism is presented, based on which the minimization problem for one-body systems is designed and analyzed.

\subsection{Wigner formalism}\label{subsec:wigner}
Given the time-independent Schr\"odinger equation,
\begin{equation}
	\left(-\frac{\nabla^2}{2}+V(\mathbf{x})\right)\psi_j(\mathbf{x})
	=E_j\psi_j(\mathbf{x}),
	\qquad
	j=0,1,2,\dots,
\end{equation}
where $\psi_j$ is the $j$-th eigenfunction with eigenvalue $E_j$ and forms an orthonormal set whose eigenvalues are in increasing order. We have the density matrix,
\begin{equation}
	\rho(\mathbf{x},\mathbf{x}')
	=\sum\limits_{j=0}^\infty P_j\psi_j(\mathbf{x})\psi_j^*(\mathbf{x}'),
\end{equation}
here $P_j$ is the probability of obtaining the $j$-th eigenstate. The Wigner function $f(\mathbf{x},\mathbf{p})\in\mathbb{R}^{2D}$ is defined by applying the Weyl transform to the density matrix,
\begin{equation}
	f(\mathbf{x},\mathbf{p})
	:=\frac{1}{(2\pi)^D}\int_{\mathbb{R}^D}\rho\left(\mathbf{x}+\frac{\mathbf{y}}{2},\mathbf{x}-\frac{\mathbf{y}}{2}\right)\exp(-i\mathbf{p}\cdot\mathbf{y})d\mathbf{y},
\end{equation}
where $D$ is the dimensionality of the system, $i=\sqrt{-1}$ represents the imaginary unit. Following the basic property of the Weyl transform, we have the expression of the density and energy as follows,
\begin{linenomath}\begin{align}\label{eqn:rho}
	\rho(\mathbf{x})
	&=\int_{\mathbb{R}^D}f(\mathbf{x},\mathbf{p})d\mathbf{p},\\
	E
	&=\iint_{\mathbb{R}^D\times\mathbb{R}^D}\left(\frac{|\mathbf{p}|^2}{2}+V\right)f(\mathbf{x},\mathbf{p})d\mathbf{x}d\mathbf{p},
\end{align}\end{linenomath}
where $V$ is the prescribed potential function. Then we have the normalization condition for the Wigner function
\begin{equation}\label{eqn:normal cond}
	\iint_{\mathbb{R}^D\times\mathbb{R}^D}f(\mathbf{x},\mathbf{p})d\mathbf{x}d\mathbf{p}
	=1.
\end{equation}

Based on the time-independent Schr\"odinger equation, one can derive the following eigenvalue problem \cite{hillery1984distribution,zhan2022wigner},
\begin{equation}\label{eqn:Hw}
	H_{\rm w}f(\mathbf{x},\mathbf{p})
	:=\frac{1}{2}\left(-\frac{\nabla_\mathbf{x}^2}{4}f(\mathbf{x},\mathbf{p})+|\mathbf{p}|^2f(\mathbf{x},\mathbf{p})+\int_{\mathbb{R}^D}V_{\rm eig}(\mathbf{x},\mathbf{p}')f(\mathbf{x},\mathbf{p}-\mathbf{p}')d\mathbf{p}'\right)
	=Ef(\mathbf{x},\mathbf{p}),\\
\end{equation}
where
\begin{linenomath}\begin{align}\label{eqn:Veig}
	V_{\rm eig}(\mathbf{x},\mathbf{p})
	&=\frac{1}{(2\pi)^D}\int_{\mathbb{R}^D}S_V(\mathbf{x},\mathbf{y})\exp(-i\mathbf{p}\cdot\mathbf{y})d\mathbf{y},\\\label{eqn:SV}
	S_V(\mathbf{x},\mathbf{y})
	&=V\left(\mathbf{x}+\frac{\mathbf{y}}{2}\right)+V\left(\mathbf{x}-\frac{\mathbf{y}}{2}\right).
\end{align}\end{linenomath}

\modify{
Although the ground state in Wigner formalism can be obtained by directly solving (\ref{eqn:Hw}), the problem dimensionality actually doubles. This issue becomes even more severe when addressing the ground state of many-body systems. Therefore, it is preferable to determine the many-body ground state through energy optimization, which focuses the computation on only single Wigner function, leveraging the descriptive capability of the Wigner formalism. In the next subsection, the energy functional for one-body systems in Wigner formalism is proposed, which is a crucial component for developing the Wigner gradient flow model.
}

\subsection{Energy functional for one-body systems in Wigner formalism}\label{subsec:energy functional}
It is noted that the basic property of the Weyl transform implies that
\begin{equation}
	\iint_{\mathbb{R}^D\times\mathbb{R}^D}|f(\mathbf{x},\mathbf{p})|^2d\mathbf{x}d\mathbf{p}
	=\iint_{\mathbb{R}^D\times\mathbb{R}^D}f(\mathbf{x},\mathbf{p})^2d\mathbf{x}d\mathbf{p}
	=(2\pi)^{-D}.
\end{equation}
Since $H_{\rm w}$ in Eq. (\ref{eqn:Hw}) is Hermitian, we can intuitively define the energy functional as
\begin{equation}\label{eqn:E*}
	E_{\rm one}[f]
	:=(2\pi)^D\iint_{\mathbb{R}^D\times\mathbb{R}^D}f(\mathbf{x},\mathbf{p})^*H_{\rm w}f(\mathbf{x},\mathbf{p}).
\end{equation}
\modify{
Owing to the variational principle, the ground state Wigner function can be obtained by minimizing $E_{\rm one}[f]$ in (\ref{eqn:E*}). To demonstrate the validity of this process, we have the following theorem.
}
\begin{theorem}
	For one-body system, suppose the validity of the change of variables $\mathbf{u}=\mathbf{x}+\mathbf{y}/2$, $\mathbf{v}=\mathbf{x}-\mathbf{y}/2$, the ground state Wigner function and the corresponding energy can be obtained by minimizing $E_{\rm one}[f]$ over real-valued functions in $L_2(\mathbb{R}^{2D})$ with the constraint in Eq. (\ref{eqn:normal cond}).
\end{theorem}
\begin{proof}
	Let $g:\mathbb{R}^{2D}\rightarrow\mathbb{R}$, $g\in L_2(\mathbb{R}^{2D})$, it follows the hypothesis that there exists expansion
	\begin{linenomath}\begin{align}\label{eqn:g expand}
		&g(\mathbf{x},\mathbf{p})
		=\sum\limits_{j,k=0}^\infty c_{jk}f_{jk}(\mathbf{x},\mathbf{p}),\\
		&f_{jk}(\mathbf{x},\mathbf{p})
		=\frac{1}{(2\pi)^D}\int_{\mathbb{R}^D}\psi_k^*\left(\mathbf{x}-\frac{\mathbf{y}}{2}\right)\psi_j\left(\mathbf{x}+\frac{\mathbf{y}}{2}\right)\exp(-i\mathbf{p}\cdot\mathbf{y})d\mathbf{y},\\
		&c_{jk}
		=(g,f_{jk})/(f_{jk},f_{jk}),
	\end{align}\end{linenomath}
	where $(\cdot,\cdot)$ is the inner produce in $L_2(\mathbb{R}^{2D})$.
	
	Particularly, $f_{jj}$ recovers the Wigner function corresponds to the $j$-th eigenstate, and
	\begin{linenomath}\begin{align}
		&(f_{jk},f_{j'k'})
		=(2\pi)^{-D}\delta_{jj'}\delta_{kk'},\\
		&H_{\rm w}f_{jk}
		=\frac{E_j+E_k}{2}f_{jk},\\
		&\iint_{\mathbb{R}^D\times\mathbb{R}^D}f_{jk}
		=\delta_{jk},\\
		&\iint_{\mathbb{R}^D\times\mathbb{R}^D}\left(\frac{|\mathbf{p}|^2}{2}+V(\mathbf{x})\right)f_{jj}
		=E_j,\label{eqn:energy fjj}
	\end{align}\end{linenomath}
	where $\delta_{jk}$ is the Kronecker delta symbol.
	
	One can deduce from the normalization condition Eq. (\ref{eqn:normal cond}) that
	\begin{equation}
		\sum\limits_{j=0}^\infty|c_{jj}|^2=1.
	\end{equation}
	Finally, by substitution of Eq. (\ref{eqn:g expand}) into Eq. (\ref{eqn:E*}) we find
	\begin{equation}
		E_{\rm one}[g]
		=\sum\limits_{j,k=0}^\infty\frac{E_j+E_k}{2}|c_{jk}|^2
		=\sum\limits_{j=0}^\infty E_j|c_{jj}|^2+\sum\limits_{0\le j<k}\frac{E_j+E_k}{2}(|c_{jk}|^2+|c_{kj}|^2)
		\ge E_0,
	\end{equation}
	with the equality holds if and only if $c_{jk}=1$ when $j=k=0$, and $c_{jk}=0$ otherwise. Moreover, it follows Eq. (\ref{eqn:energy fjj}) that $E_{\rm one}[f_{00}]=E_0$ in this situation.
\end{proof}

\modify{
Based on the above theorem, two approaches can be considered for the ground state calculations in Wigner formalism, i.e., optimization approach for minimizing the energy functional, and solving the equation derived from the first-order necessary condition. In our previous work \cite{zhan2022wigner}, the latter approach has been explored, in which the ground state of a one-dimensional two-body system is computed. However, extending this method to three-dimensional cases is a nontrivial task. Moreover, applying the method to more complicated system requires calculating more Wigner functions corresponding to different orbitals, greatly increasing computational demands. In the next section, a gradient flow model for ground state calculations in Wigner formalism of many-body systems is presented, fully utilizing the depicting capability of the Wigner formalism and reducing the computational burden associated with simulating numerous Wigner functions.
}

\section{Gradient flow model for ground state Wigner function}\label{sec:dft}
\modify{
Utilizing the result in Section \ref{subsec:energy functional}, one can derive the Wigner gradient flow model for one-body systems over real-valued functions in $L_2(\mathbb{R}^D\times\mathbb{R}^D)$ as
\begin{equation}
	\left\{\begin{array}{cl}
		\displaystyle\frac{\partial}{\partial t}f(\mathbf{x},\mathbf{p},t)
		=-\frac{\delta E_{\rm one}[f]}{\delta f(\mathbf{x},\mathbf{p},t)}
		=-2(2\pi)^DH_{\rm w}f(\mathbf{x},\mathbf{p},t),
		&t>0,\\
		\displaystyle\iint_{\mathbb{R}^D\times\mathbb{R}^D}f(\mathbf{x},\mathbf{p},t)d\mathbf{x}d\mathbf{p}
		=1,
		&t\ge0.
	\end{array}\right.
\end{equation}
In this section, the Kohn-Sham model is introduced to resolve the high dimensionality of many-body problems. Following a similar process as the last section, the energy functional of many-body systems in Wigner formalism is presented, based on which a Wigner gradient flow model will be developed.
}

\subsection{Kohn-Sham model}\label{subsec:KS}
Within the context of density functional theory, the many-body wave function can be approximated by the Slater determinant of the Kohn-Sham orbital functions, which satisfy the following Kohn-Sham equations,
\begin{equation}
	\left(-\frac{\nabla^2}{2}+V_{\rm KS}[\rho](\mathbf{x})\right)\psi^{\rm KS}_j(\mathbf{x})=\varepsilon_j\psi_j^{\rm KS}(\mathbf{x}),
	\qquad
	j=1,2,\dots,
\end{equation}
where
\begin{equation}
	\rho(\mathbf{x})
	=\sum\limits_{j=1}^{N_o}P_j^{\rm KS}\rho_j^{\rm KS}(\mathbf{x}),
	\qquad
	\rho_j^{\rm KS}(\mathbf{x})
	=|\psi_j^{\rm KS}(\mathbf{x})|^2,
\end{equation}
$N_o$ is the number of occupied orbitals, $\psi_j^{\rm KS}$ are the $j$-th eigenstate with eigenvalue $\varepsilon_j$ in increasing order, and $P_j^{\rm KS}$ is the occupation number of $\psi_j^{\rm KS}$. In particular,
\begin{equation}
	V_{\rm KS}[\rho](\mathbf{x})
	=V_{\rm ext}(\mathbf{x})+U_{\rm H}[\rho](\mathbf{x})+V_{\rm xc}[\rho](\mathbf{x}).
\end{equation}
Here $V_{\rm ext}$ stands for the external potential. The second term depicts the electron-electron interaction subject to the interaction function $V_{\rm ee}$, which can be attained following the variational principle of the interaction energy,
\begin{equation}
	U_{\rm H}[\rho]
	=\frac{1}{2}\iint_{\mathbb{R}^D\times\mathbb{R}^D}V_{\rm ee}(|\mathbf{x}-\mathbf{x}'|)\rho(\mathbf{x})\rho(\mathbf{x}')d\mathbf{x}d\mathbf{x}',
\end{equation}
as
\begin{equation}
	V_{\rm H}[\rho](\mathbf{x})
	=\frac{\delta U_{\rm H}[\rho]}{\delta\rho(\mathbf{x})}
	=\int_{\mathbb{R}^D} V_{\rm ee}(|\mathbf{x}-\mathbf{x}'|)\rho(\mathbf{x}')d\mathbf{x}'.
\end{equation}
And the last term
\begin{equation}
	V_{\rm xc}[\rho](\mathbf{x})
	=\frac{\delta E_{\rm xc}[\rho]}{\rho(\mathbf{x})}
\end{equation}
is the exchange-correlation potential. Since an analytic expression of $V_{\rm xc}$ is often unknown in the most situations, approximations are required for the this term. 

Subsequently, the descriptive capability of the Wigner formalism can be presented in a straightforward way in the sense that single mixed state Wigner function represents the whole DFT system,
\begin{equation}
	f_{\rm gs}^{\rm KS}(\mathbf{x},\mathbf{p})
	=\frac{1}{(2\pi)^D}\int_{\mathbb{R}^D}\rho_{\rm gs}^{\rm KS}\left(\mathbf{x}+\frac{\mathbf{y}}{2},\mathbf{x}-\frac{\mathbf{y}}{2}\right)\exp(-i\mathbf{p}\cdot\mathbf{y})d\mathbf{y},
\end{equation}
where
\begin{equation}
	\rho_{\rm gs}^{\rm KS}(\mathbf{x},\mathbf{x}')
	=\sum\limits_{j=1}^{N_o}P_j^{\rm KS}\psi_j^{\rm KS}(\mathbf{x})(\psi_j^{\rm KS})^*(\mathbf{x}').
\end{equation}

For brevity, we focus on the closed-shell system in the following, i.e., taking $P_j^{\rm KS}=P_{\rm occ}^{\rm cs}=2$ for all $1\le j\le N_o$.

\subsection{A gradient flow model for many-body systems in Wigner formalism}\label{subsec:KS gradient flow}
To fully utilize the advantages of Wigner formalism that single function depicts the whole system, it is natural to consider energy minimization approaches for ground state calculations in Wigner formalism for many-body systems. In particular, we have the following energy functional,
\begin{equation}
	E^{\rm KS}[\rho]
	=\sum\limits_{j=1}^{N_o}P_j\varepsilon_j-U_{\rm H}[\rho]-\int_{\mathbb{R}^D} V_{\rm xc}[\rho]\rho+E_{\rm xc}[\rho].
\end{equation}
The corresponding minimization problem is written as
\begin{equation}
	\left\{\begin{array}{lc}
		\min	&E^{\rm KS}[\rho],\\
		\text{s.t.}	&\displaystyle\int_{\mathbb{R}^D}\rho=N_e.
	\end{array}\right.
\end{equation}
where $N_e$ is the electron number. 
The Wigner-Weyl of the energy operator gives
\begin{equation}\label{eqn:Hw KS}
	H_{\rm w}^{\rm KS}[\rho]f(\mathbf{x},\mathbf{p})
	=\frac{1}{2}\left(-\frac{\nabla_\mathbf{x}^2}{4}f(\mathbf{x},\mathbf{p})+|\mathbf{p}|^2f(\mathbf{x},\mathbf{p})+\int_{\mathbb{R}^D}V_{\rm eig}^{\rm KS}[\rho](\mathbf{x},\mathbf{p}')f(\mathbf{x},\mathbf{p}-\mathbf{p}')d\mathbf{p}'\right),
\end{equation}
where the last term corresponds to the one in Eq. (\ref{eqn:Hw}) with the potential function in Eq. (\ref{eqn:SV}) replaced by $V_{\rm KS}[\rho]$. Then the energy functional of the Kohn-Sham model is represented as
\begin{equation}
	E_{\rm w}^{\rm KS}[f]
	=\frac{(2\pi)^D}{P_{occ}^{cs}}\iint_{\mathbb{R}^D\times\mathbb{R}^D}f(\mathbf{x},\mathbf{p})^*H_{\rm w}^{\rm KS}[\rho]f(\mathbf{x},\mathbf{p})d\mathbf{x}d\mathbf{p}
	-E_{\rm ee}[\rho]-\int_{\mathbb{R}^D}V_{\rm xc}[\rho]\rho+E_{\rm xc}[\rho],
\end{equation}
with the density given by Eq. (\ref{eqn:rho}). Additionally, the constraint becomes
\begin{equation}
	\iint_{\mathbb{R}^D\times\mathbb{R}^D}f(\mathbf{x},\mathbf{p})d\mathbf{x}d\mathbf{p}
	=N_e.
\end{equation}
Suppose the second-order continuity of $V_{\rm xc}[\rho]$ w.r.t. $\rho$, one can derive the following results by calculations for the Wigner function corresponding to the ground state of the Kohn-Sham model,
\begin{equation}
	\begin{aligned}
		\frac{\delta E_{\rm w}^{\rm KS}[f]}{\delta f(\mathbf{x},\mathbf{p})}
		&=(2\pi)^DH_{\rm w}^{\rm KS}[\rho]f(\mathbf{x},\mathbf{p})
		+\left(\frac{\delta U_{\rm H}[\rho]}{\delta\rho(x)}+\frac{\delta V_{\rm xc}[\rho]}{\delta\rho(x)}\right)\rho[f](x)\\
		&\quad-\frac{\delta U_{\rm H}[\rho]}{\rho[f](\mathbf{x})}\frac{\delta \rho[f](\mathbf{x})}{\delta f(\mathbf{x},\mathbf{p})}
		-\left(\frac{\delta V_{\rm xc}[\rho]}{\delta \rho[f](\mathbf{x})}\rho[f](\mathbf{x})+V_{\rm xc}[\rho](\mathbf{x})\right)\frac{\delta\rho[f](\mathbf{x})}{\delta f(\mathbf{x},\mathbf{p})}
		+\frac{\delta E_{\rm xc}[\rho]}{\delta \rho[f](\mathbf{x})}\frac{\delta\rho[f](\mathbf{x})}{\delta f(\mathbf{x},\mathbf{p})}\\
		&=(2\pi)^DH_{\rm w}^{\rm KS}[\rho]f(\mathbf{x},\mathbf{p}).
	\end{aligned}
\end{equation}
Therefore, the Wigner gradient flow model for the Kohn-Sham model shares the form as one-body systems,
\begin{equation}\label{eqn:wigner dft gradient flow}
	\left\{\begin{array}{cl}
		\displaystyle\frac{\partial}{\partial t}f(\mathbf{x},\mathbf{p},t)=-(2\pi)^DH_{\rm w}^{\rm KS}[\rho]f(\mathbf{x},\mathbf{p},t),
		&t>0,\\
		\displaystyle\iint_{\mathbb{R}^D\times\mathbb{R}^D}f(\mathbf{x},\mathbf{p},t)d\mathbf{x}d\mathbf{p}=N_e,
		&t\ge0.
	\end{array}\right.
\end{equation}

\section{Numerical scheme}\label{sec:numerical method}
Despite the advantage of the Wigner formalism that allows the depiction of the entire system by a single function, the problem dimensionality doubles compared to the original formulation. Therefore, efficient numerical scheme and discretization are crucial for ground state calculations in Wigner formalism. In Sect. \ref{subsec:splitting}, an operator splitting scheme is introduced, which validates the parallel implementations of the gradient flow model. Moreover, leveraging the structure of convolution operator in Eq. (\ref{eqn:Hw KS}), the Fourier pseudo-spectral method is adopted for discretization in Sect. \ref{subsec:fourier}. Consequently, an efficient parallelizable algorithm is obtained, with the computational complexity $O(n_{\rm DoF}\log n_{\rm DoF})$.

\subsection{Operator splitting method}\label{subsec:splitting}
With three operators
\begin{linenomath}\begin{align}
	\mathcal{A}f(\mathbf{x},\mathbf{p},t)
	&:=-\frac{\nabla_\mathbf{x}^2}{4}f(\mathbf{x},\mathbf{p},t),\\
	\mathcal{B}f(\mathbf{x},\mathbf{p},t)
	&:=|\mathbf{p}|^2f(\mathbf{x},\mathbf{p},t),\\
	\mathcal{C}[\rho]f(\mathbf{x},\mathbf{p},t)
	&:=\int_{\mathbb{R}^D}V_{\rm eig}^{\rm KS}[\rho](\mathbf{x},\mathbf{p}')f(\mathbf{x},\mathbf{p}-\mathbf{p}')d\mathbf{p}',
\end{align}\end{linenomath}
the evolution equation in (\ref{eqn:wigner dft gradient flow}) can be rewritten as
\begin{equation}
	\frac{\partial}{\partial t}f(\mathbf{x},\mathbf{p},t)
	=-(\mathcal{A}+\mathcal{B}+\mathcal{C})f(\mathbf{x},\mathbf{p},t),
\end{equation}
which can be split into the following three sub-equations in an alternating manner
\begin{equation}\label{eqn:split system}
	\left\{\begin{array}{ll}
		\text{(A)}
		&\displaystyle\frac{\partial}{\partial t}f(\mathbf{x},\mathbf{p},t)
		=-\mathcal{A}f(\mathbf{x},\mathbf{p},t)
		=\frac{\nabla_\mathbf{x}^2}{4}f(\mathbf{x},\mathbf{p},t),\\
		\text{(B)}
		&\displaystyle\frac{\partial}{\partial t}f(\mathbf{x},\mathbf{p},t)
		=-\mathcal{B}f(\mathbf{x},\mathbf{p},t)
		=-|\mathbf{p}|^2f(\mathbf{x},\mathbf{p},t),\\
		\text{(C)}
		&\displaystyle\frac{\partial}{\partial t}f(\mathbf{x},\mathbf{p},t)
		=-\mathcal{C}[\rho]f(\mathbf{x},\mathbf{p},t)
		=-\int_{\mathbb{R}^D}V_{\rm eig}^{\rm KS}[\rho](\mathbf{x},\mathbf{p}')f(\mathbf{x},\mathbf{p}-\mathbf{p}')d\mathbf{p}'.
	\end{array}\right.
\end{equation}
We employ a simple linearization and the well-known second-order Strang method \cite{strang1968construction},
\begin{equation}
	f(\mathbf{x},\mathbf{p},t^{n+1})
	=e^{-\mathcal{A}\Delta t/2}e^{-\mathcal{B}\Delta t/2}e^{-\mathcal{C}[c_{\rm nor}[\rho^*]\rho^*]\Delta t}e^{-\mathcal{B}\Delta t/2}e^{-\mathcal{A}\Delta t/2}f(\mathbf{x},\mathbf{p},t^n)+O(\Delta t^2),
\end{equation}
where $t^n$ and $t^{n+1}$ are two adjacent discrete moment, $\Delta t=t^{n+1}-t^n$,
\begin{equation}
	\rho^*(\mathbf{x})
	=\int_{\mathbb{R}^D}e^{-\mathcal{B}\Delta t/2}e^{-\mathcal{A}\Delta t/2}f(\mathbf{x},\mathbf{p},t^n)d\mathbf{p},
	\qquad
	c_{\rm nor}[\rho]
	=\left(\int_{\mathbb{R}^D}\rho(\mathbf{x})d\mathbf{x}\right)^{-1}\cdot N_{\rm e}.
\end{equation}

\subsection{Fourier spectral collocation method}\label{subsec:fourier}
\modify{
It is worth mentioning that variables decouple when solving sub-equation (A) and sub-equation (B) in Eq. (\ref{eqn:split system}), facilitating parallel implementation. Furthermore, the solution of sub-equation (C) actually corresponds to the independent evolution of Fourier coefficients, making the Fourier spectral collocation method an ideal choice for discretization. More specifically, we discretize each direction separately: (i) the $\mathbf{x}$-domain is either truncated or chosen as $\mathcal{X}=\prod_{j=1}^D[0,a_j]$, admitting a zero Dirichlet boundary condition or a periodic boundary condition; (ii) given the decay property of the Wigner function when $|\mathbf{p}|\rightarrow+\infty$, a simple nullification is adopted outside a sufficiently large $\mathbf{p}$-domain $\mathcal{P}=[-L/2,L/2]^D$. Consequently, the function $f(\mathbf{x},\mathbf{p},t)$ can be represented as
}
\begin{equation}\label{eqn:f expand}
	f(\mathbf{x},\mathbf{p},t)
	=\sum\limits_{\nu_p \in\mathcal{I}_p}c_{\nu_p}(\mathbf{x},t)\phi_{\nu_p}(\mathbf{p})
	=\sum\limits_{\nu_p \in\mathcal{I}_p}\sum\limits_{\nu_x \in\mathcal{I}_x}c_{\nu_p,\nu_x}(t)\psi_{\nu_x}(\mathbf{x})\phi_{\nu_p}(\mathbf{p}),
\end{equation}
where $x_j$, $p_j$, $(\nu_x)_j$, $(\nu_p)_j$ and $(N_x)_j$ are the $j$-th components of $\mathbf{x}$, $\mathbf{p}$, $\nu_x$, $\nu_p$, and $N_x$ respectively,  
\begin{linenomath}\begin{align}
	\psi_{\nu_x}(\mathbf{x})
	&=\exp\left(\sum\limits_{j=1}^D2\pi i(\nu_x)_jx_j/a_j\right),\\
	\phi_{\nu_p}(\mathbf{p})
	&=\exp\left(\sum\limits_{j=1}^D2\pi i(\nu_p)_j(p_j-L/2)/L\right),\\
	\mathcal{I}_x
	&:=\{\nu\in\mathbb{Z}^D:-(N_x)_j/2\le \nu_j<(N_x)_j/2,1\le j\le D\},\\
	\mathcal{I}_p
	&:=\{\nu\in\mathbb{Z}^D:-N_p/2\le \nu_j<N_p/2,1\le j\le D\},
\end{align}\end{linenomath}
$N_x$ and $N_p$ stand for truncation orders in $\mathbf{x}$ and $\mathbf{p}$ direction, respectively.
Then interpolation of the Wigner function into the solution space
\begin{equation}
	\{\psi_{\nu_x}(\mathbf{x})\phi_{\nu_p}(\mathbf{p}):\nu_x \in\mathcal{I}_x,\nu_p \in\mathcal{I}_p\},
\end{equation}
has coefficients as follows,
\begin{linenomath}\begin{align}
	c_{\nu_p}(\mathbf{x},t)
	&=\frac{1}{N_p^D}\sum\limits_{\mu_p \in\mathcal{J}_p}f(\mathbf{x},\mathbf{p}_{\mu_p},t)\phi_{-\nu_p}(\mathbf{p}_{\mu_p}),\\
	c_{\nu_x,\nu_p}(t)
	&=\frac{1}{\prod_{j=1}^D(N_x)_j}\sum\limits_{\mu_x \in\mathcal{J}_x}c_{\nu_p}(\mathbf{x}_{\mu_x},t)\psi_{-\nu_x}(\mathbf{x}_{\mu_x}),
\end{align}\end{linenomath}
where $(\mu_x)_j$, $(\mu_p)_j$, $(\mu_x)_ja_j/(N_x)_j$ and $(\mu_p)_jL/N_p-L/2$ are the $j$-th component of $\mu_x$, $\mu_p$, $\mathbf{x}_{\mu_x}$ and $\mathbf{p}_{\mu_p}$, respectively,
\begin{linenomath}\begin{align}
	\mathcal{J}_x
	&:=\{\mu_x\in\mathbb{Z}^D:0\le(\mu_x)_j<(N_x)_j,1\le j\le D\},\\
	\mathcal{J}_p
	&:=\{\mu_p\in\mathbb{Z}^D:0\le(\mu_p)_j<N_p,1\le j\le D\}.
\end{align}\end{linenomath}
\begin{remark}
	The degree of freedoms for the proposed method can be interpreted as the following equivalent sets in the sense of discrete Fourier transform and inverse discrete Fourier transform:
	\begin{equation}\label{eqn:equivalent set}
		\{f(\mathbf{x}_{\mu_x},\mathbf{p}_{\mu_p},t)\}_{\mu_x \in\mathcal{J}_x,\mu_p \in\mathcal{J}_p},
		\qquad
		\{c_{\nu_p}(\mathbf{x}_{\mu_x},t)\}_{\mu_x \in\mathcal{J}_x,\nu_p \in\mathcal{I}_p},
		\qquad
		\{c_{\nu_x,\nu_p}(t)\}_{\nu_x \in\mathcal{I}_x,\nu_p \in\mathcal{I}_p}.
	\end{equation}
	Consequently, the evolution of sub-equations can be applied to one the these equivalent sets, with only one index involved and another one decouples. Therefore, the evolution of the gradient flow model can be efficiently accomplished by parallel implementations.
\end{remark}
Now we consider the evolution for sub-equations separately,
\begin{itemize}
	\item Sub-equation (A): Direct substitution of Eq. (\ref{eqn:f expand}) and the orthogonality of $\{\phi_{\nu_p}\}_{\nu_p \in\mathcal{I}_p}$ imply that
	\begin{equation}
		\frac{\partial}{\partial t}c_{\nu_p}(\mathbf{x},t)
		=\frac{\nabla_\mathbf{x}^2}{4}c_{\mu_p}(\mathbf{x},t),
		\qquad
		\mu_p \in\mathcal{I}_p,
	\end{equation}
	which is equivalent to
	\begin{equation}
		\frac{d}{dt}c_{\mu_x,\mu_p}(t)
		=-\pi^2\sum\limits_{j=1}^D\left(\frac{(\nu_x)_j}{a_j}\right)^2c_{\mu_x,\mu_p}(t),
		\qquad
		\mu_x \in\mathcal{I}_x,\mu_p \in\mathcal{I}_p.
	\end{equation}
	Therefore the numerical solution can be explicitly given as
	\begin{equation}\label{eqn:sol A}
		c_{\mu_x,\mu_p}(t^{n+1})
		=\exp\left(-\pi^2\Delta t\sum\limits_{j=1}^D\left(\frac{(\nu_x)_j}{a_j}\right)^2\right)c_{\mu_x,\mu_p}(t^n),
		\qquad
		\mu_x \in\mathcal{I}_x,\mu_p \in\mathcal{I}_p.
	\end{equation}
	\item Sub-equation (B): We consider the effect acts on the first set of (\ref{eqn:equivalent set}),
	\begin{equation}
		\frac{\partial}{\partial t}f(\mathbf{x}_{\mu_x},\mathbf{p},t)
		=-|\mathbf{p}|^2f(\mathbf{x}_{\mu_x},\mathbf{p},t),
		\qquad
		\mu_x \in\mathcal{J}_x,
	\end{equation}
	whose solution is
	\begin{equation}\label{eqn:sol B}
		f(\mathbf{x}_{\mu_x},\mathbf{p}_{\mu_p},t^{n+1})
		=\exp(-|\mathbf{p}_{\mu_p}|^2\Delta t)f(\mathbf{x}_{\mu_x},\mathbf{p}_{\mu_p},t^n),
		\qquad
		\mu_x \in\mathcal{J}_x,\mu_p \in\mathcal{J}_p.
	\end{equation}
	\item Sub-equation (C): Suppose $V_{\rm eig}^{\rm KS}[\rho](\mathbf{x},\mathbf{p})$ decay to zero outside some $\mathbf{p}$-domain, it follows Poisson summation formula that
	\begin{equation}
		\begin{aligned}
			\mathcal{C}[\rho]f(\mathbf{x},\mathbf{p},t)
			&\approx\left(\frac{\Delta y}{2\pi}\right)^D\sum\limits_{\mu\in\mathbb{Z}^3}S_{V_{\rm KS}[\rho]}(\mathbf{x},\mathbf{y}_\mu)\int_\mathcal{P}\exp(-i\mathbf{p}'\cdot\mathbf{y})\sum\limits_{\nu_p \in\mathcal{I}_p}c_{\nu_p}(\mathbf{x},t)\phi_{\nu_p}(\mathbf{p})d\mathbf{p}\\
			&=\sum\limits_{\nu_p \in\mathcal{I}_p}S_{V_{\rm KS}[\rho]}(\mathbf{x},\mathbf{y}_{-\nu_p})c_{\nu_p}(\mathbf{x},t)\phi_{\nu_p}(\mathbf{p}),
		\end{aligned}
	\end{equation}
	where it requires that $L\cdot\Delta y=2\pi$ \cite{xiong2016advective}, and $\mathbf{y}_{\nu_p}$ has $j$-th component $(\nu_p)_j\Delta y$. Therefore, the sub-equation (C) can be converted to
	\begin{equation}
		\frac{\partial}{\partial t}c_{\nu_p}(\mathbf{x},t)
		=S_{V_{\rm KS}[\rho]}(\mathbf{x},\mathbf{y}_{-\nu_p})c_{\nu_p}(\mathbf{x},t),
		\qquad
		\nu_p \in\mathcal{I}_p.
	\end{equation}
	With simple linearization, one can obtain an explicit expression of the solution as
	\begin{equation}\label{eqn:sol C}
		c_{\nu_p}(\mathbf{x}_{\mu_x},t^{n+1})
		=\exp\Big(-\Delta tS_{V_{\rm KS}[c_{\rm nor}c_0^n]}(\mathbf{x}_{\mu_x},\mathbf{y}_{-\nu_p})\Big)c_{\nu_p}(\mathbf{x}_{\mu_x},t^n),
		\qquad
		\mu_x \in\mathcal{J}_x,\nu_p \in\mathcal{I}_p,
	\end{equation}
	where $c_{\rm nor}$ stands for the normalization coefficient such that $\int_\mathcal{X}c_{\rm nor}c_0$ recovers the electron number.
\end{itemize}
Moreover, the concerned quantities, i.e., density and energy can be expressed as
\begin{linenomath}\begin{align}\label{eqn:rho-discrete}
	\rho(\mathbf{x},t)
	&=L^Dc_0(\mathbf{x},t),\\
	\label{eqn:E-discrete}
	E(t)
	&=\frac{L^{D-1}}{2}\sum\limits_{\nu_p \in\mathcal{I}_p^0}\frac{L^3}{2\pi^2|\nu_p|^2+12\delta_{0,|\nu_p|}/D}|\mathcal{X}|c_{0,\nu_p}(t),
\end{align}\end{linenomath}
with
\begin{equation}
	\mathcal{I}_p^0
	:=\{\nu_p \in\mathcal{I}_p:\#\{j:(\nu_p)_j=0\}\ge2\}.
\end{equation}
Finally, denoting the unknowns in (\ref{eqn:equivalent set}) as $F_\mu$, $C_{\nu,\mu}$, $C_\nu$, respectively, the flowchart of the proposed numerical method for solving the DFT Wigner gradient flow model is presented in the algorithm below,
\begin{algorithm}[H]
	\caption{DFT Wigner gradient flow model solving the ground state Wigner function.}
	\label{alg:lobpcg}
	\begin{algorithmic}[1]
		\REQUIRE{Initial guess $C_{\nu,\mu}^{\rm init}$, time step size $dt $, iteration step for single test $N_{\rm test}$, maximum test step $N^{\rm max}_{\rm test}$, tolerance $Tol $.}
		\ENSURE{Fourier coefficients $C_{\nu,\mu}$ for the ground state Wigner function, ground state density $\rho_{\rm out}$ and energy $E_{\rm out}$.}
		\STATE{Set $C_{\nu,\mu}=C_{\nu,\mu}^{\rm init}$, calculate $\rho^0$ by (\ref{eqn:rho-discrete}).}
		\FOR{$n=1:N_{\rm test}^{\rm max}$}
		\FOR{$n_{\rm evolve}=1:N_{\rm test}$}
		\STATE{Attain $C_\nu$ by applying the discrete Fourier transform to $C_{\nu,\mu}$ w.r.t. variable $\mathbf{x}_\mu$, update $C_{\nu}$ following (\ref{eqn:sol A}) with time step size $dt/2$, recover $C_{\nu,\mu}$ by the inverse discrete Fourier transform.}
		\STATE{Attain $F_\mu$ by applying the discrete Fourier transform to $C_{\nu,\mu}$ w.r.t. variable $\mathbf{p}_\mu$, update $F_\mu$ following (\ref{eqn:sol B}) with time step size $dt/2$, recover $C_{\nu,\mu}$ by the inverse discrete Fourier transform.}
		\STATE{Calculate density $S_{V_{\rm KS}[c_{\rm nor}c_0^*]}(\mathbf{x}_{\mu_x},y_{\nu_p})$ with $C_{0,\mu}$, update $C_{\nu,\mu}$ following (\ref{eqn:sol C}) with time step size $dt $.}
		\STATE{Attain $F_\mu$ by applying the discrete Fourier transform to $C_{\nu,\mu}$ w.r.t. variable $\mathbf{p}_\mu$, update $F_\mu$ following (\ref{eqn:sol B}) with time step size $dt/2$, recover $C_{\nu,\mu}$ by the inverse discrete Fourier transform.}
		\STATE{Attain $C_\nu$ by applying the discrete Fourier transform to $C_{\nu,\mu}$ w.r.t. variable $\mathbf{x}_\mu$, update $C_{\nu}$ following (\ref{eqn:sol A}) with time step size $dt/2$, recover $C_{\nu,\mu}$ by the inverse discrete Fourier transform.}
		\ENDFOR
		\STATE{Calculate density $\rho^n$ and energy $E^n$ by (\ref{eqn:rho-discrete}) and (\ref{eqn:E-discrete}), evaluate error $e_\rho=\|\rho^n-\rho^{n-1}\|_2$.}
		\IF{$e_\rho<Tol$}
		\STATE{Set $\rho^{\rm out}=\rho^n$, $E^{\rm out}=E^n$.}
		\STATE{Break.}
		\ENDIF
		\ENDFOR
	\end{algorithmic}
\end{algorithm}
\begin{remark}
	It can observed that each evolution depends on only one variable, which enables the acceleration by parallel calculations according to another one.
\end{remark}
\begin{remark}
	Since the evolution of each sub-equation hs an explicit expression, the computational complexity of the Fourier transform and inverse Fourier transform dominates in the evolution of single time step. Hence, the computational cost of single time step is $O(N_{\rm DoF}\log N_{\rm DoF})$ for degree of freedom number $N_{\rm DoF}$.
\end{remark}

\section{Numerical results}\label{sec:numerical result}
In this section, two toy models are presented to validate the proposed method. Firstly, the periodic extension of one-dimensional delta-interacting Hooke's atom is examined, where a system with up to $2^{21}$ electron is tested. 
Subsequently, a three-dimensional system is considered for investigating the effect of Coulomb interaction. The three-dimensional system is generated by the periodic extension of three-dimensional Hooke's atom, subjecting to external potential and Hartree potential. 
Moreover, a system with central absence is exhibited for the further illustration of the descriptive capability of the Wigner formalism.
Anticipated spectral accuracy can be successfully observed in all the computations, while multiple-cell simulations recover the desired errors compared to single-cell simulations. These examples demonstrate the potential applications of our approach to large-scale systems and systems with defects.


In particular, the models in this section can be regarded as periodic extensions of quantum systems within a single cell subjecting to periodic boundary conditions. Consequently, the system on multiple cells can be easily obtained by extending both domain and potential setup periodically. Therefore, the DFT systems within a single cell will be introduced in the following context.

All numerical experiments are implemented using a workstation, with two AMD Epyc 7713 Processor 64 Core 2.0GHz 256MB L3 Cache (total 128 cores), 881GB memory. The operation system is Ubuntu 22.04.

\subsection{1-D delta-interacting Hooke's atom}
\begin{figure}[H]
	\centering
	\includegraphics[width=.45\linewidth]{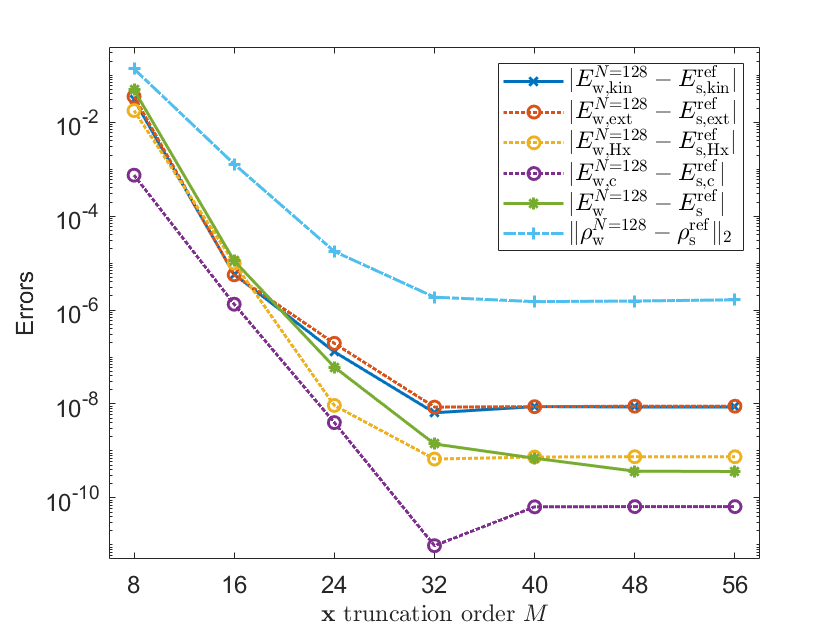}
	\includegraphics[width=.45\linewidth]{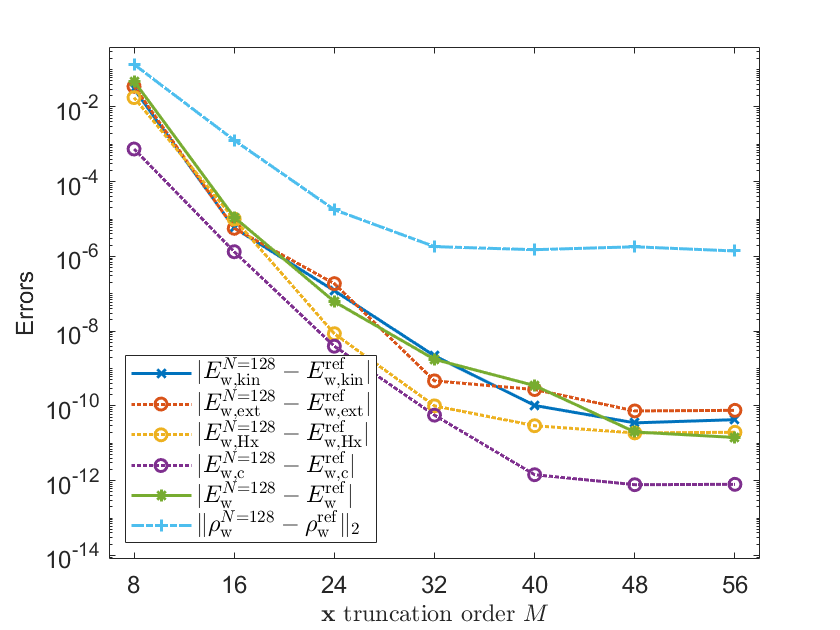}
	\caption{Decay of errors w.r.t. Schr\"odinger (left)/Wigner (left) reference for different $\mathbf{x}$ truncation order in single-cell simulations.}
	\label{fig:1d err-M 1cell}
\end{figure}

\begin{figure}[H]
	\centering
	\includegraphics[width=.45\linewidth]{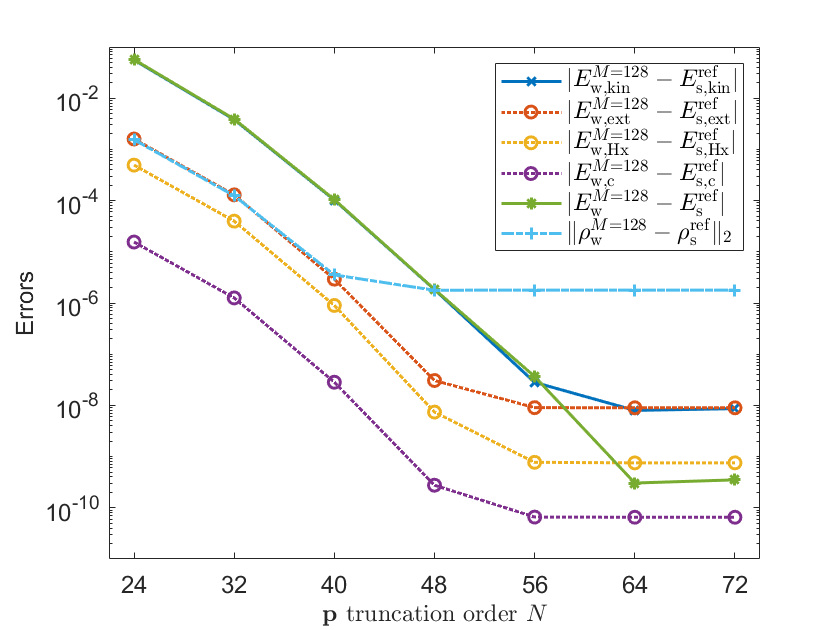}
	\includegraphics[width=.45\linewidth]{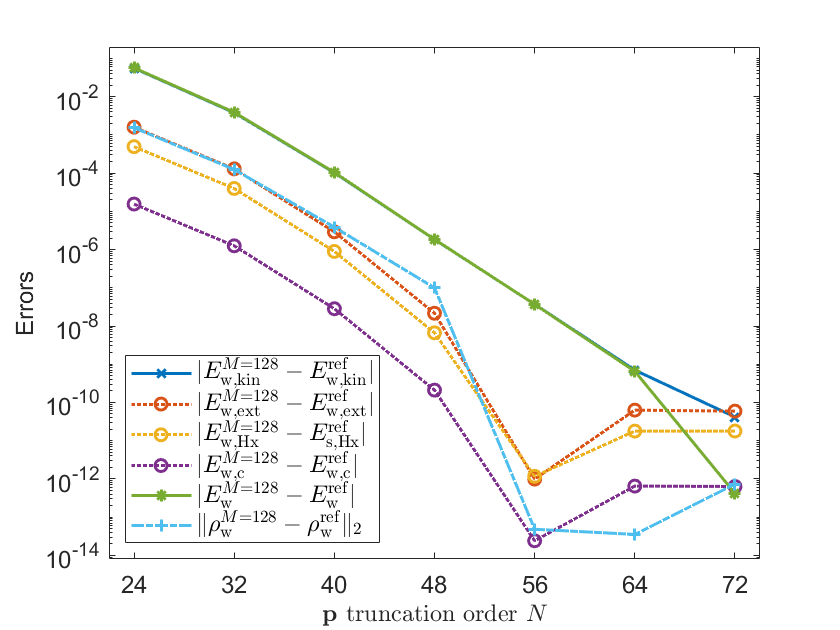}
	\caption{Decay of errors w.r.t. Schr\"odinger (left)/Wigner (left) reference for different $\mathbf{p}$ truncation order in single-cell simulations.}
	\label{fig:1d err-N 1cell}
\end{figure}

\begin{figure}[H]
	\centering
	\includegraphics[width=.48\linewidth]{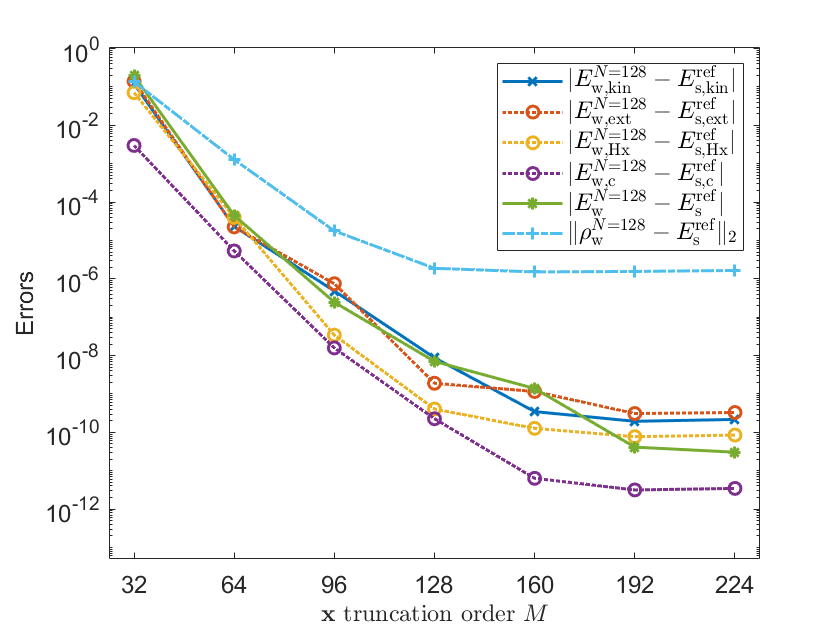}
	\includegraphics[width=.48\linewidth]{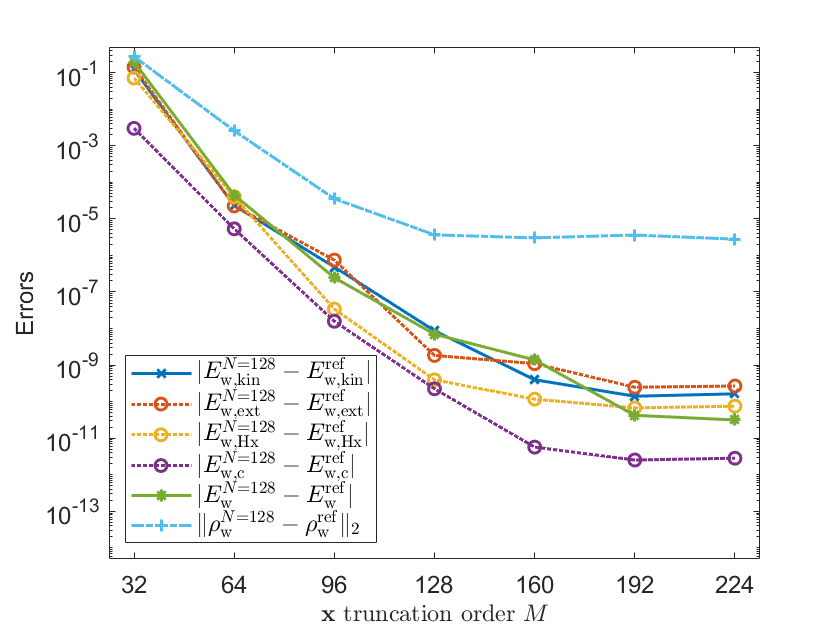}\\
	\includegraphics[width=.48\linewidth]{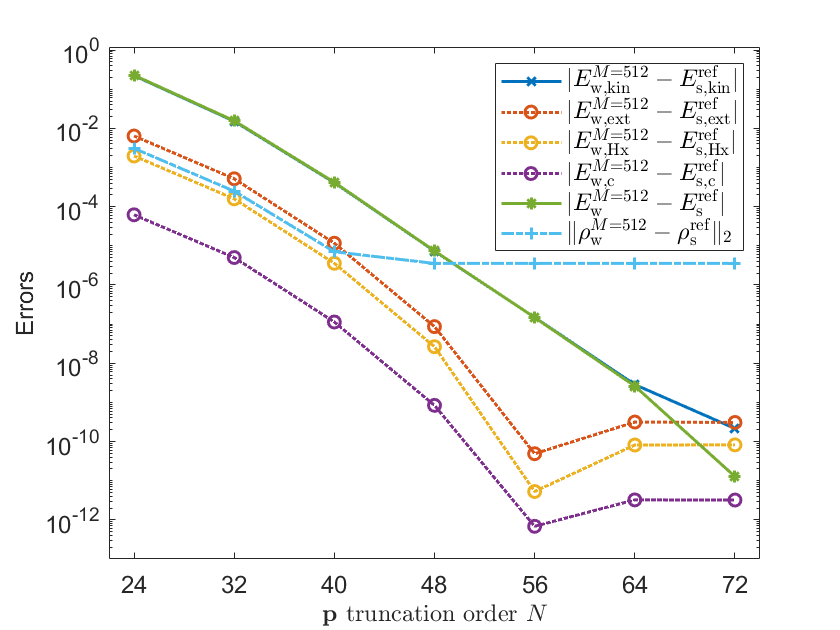}
	\includegraphics[width=.48\linewidth]{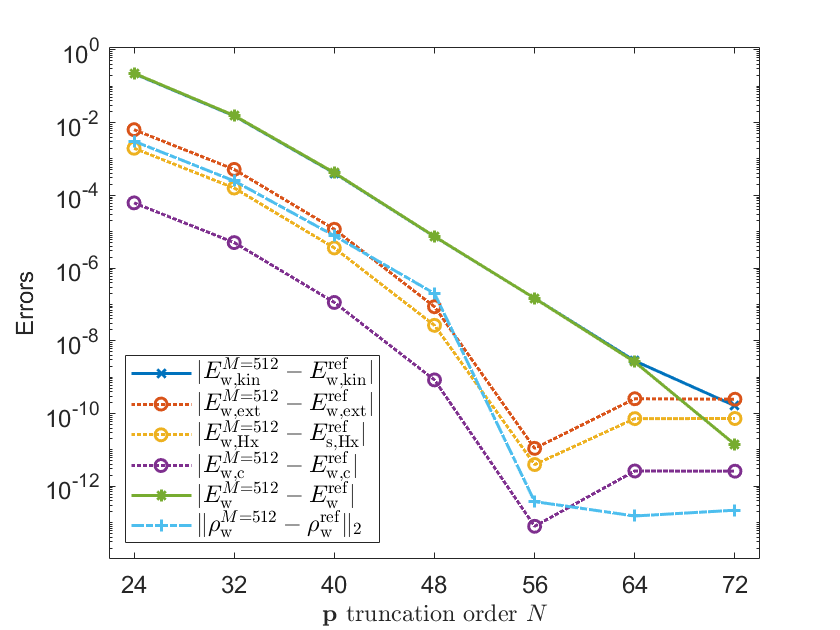}
	\caption{Decay of errors w.r.t. Schr\"odinger (left)/Wigner (left) reference for different $\mathbf{x}$ (top) $\mathbf{p}$ (bottom) truncation order in quadruple cell simulations.}
	\label{fig:1d err 4cell}
\end{figure}

In this example, simulations are conducted on both single-cell and quadruple-cell systems to validate the proposed method. The results are summarized as Fig. \ref{fig:1d err-M 1cell}, Fig. \ref{fig:1d err-N 1cell} for the single-cell simulations and Fig. \ref{fig:1d err 4cell} for the quadruple-cell simulations. Additionally, the numerical results in both the Schr\"odinger formalism and Wigner formalism with sufficient truncation order are employed as the numerical reference. Moreover, a comparison between the Wigner results in multiple-cell simulations and those in single-cell simulations is presented in Table \ref{tab:1d multi-cell}.

Consider a one-dimensional two-particle system
\begin{equation}
	\left(-\frac{1}{2}\frac{\partial}{\partial x_1}-\frac{1}{2}\frac{\partial}{\partial x_2}
	+V_{\rm ext}(x_1)+V_{\rm ext}(x_2)
	+V_{\rm ee}(x_1-x_2)\right)\Psi(x_1,x_2)
	=E\Psi(x_1,x_2),
\end{equation}
where
\begin{equation}
	V_{\rm ext}(x)=\frac{1}{2}x^2,
	\qquad
	V_{\rm ee}(x)=\delta(x).
\end{equation}
This system can be approximated by the Kohn-Sham model
\begin{equation}
	\left(-\frac{1}{2}\frac{d^2}{dx^2}+V_{KS}[\rho](x)\right)\psi_j
	=\varepsilon_j\psi_j(x),
\end{equation}
where 
\begin{equation}
	V_{\rm KS}[\rho](x)
	=V_{\rm ext}(x)+V_{\rm H}[\rho](x)+V_{\rm xc}[\rho](x),
\end{equation}
with
\begin{equation}
	V_{\rm H}[\rho](x)=\frac{\delta U[\rho]}{\delta\rho(x)},
	\qquad\text{and}\qquad
	V_{\rm xc}[\rho](x)=\frac{\delta(E_{\rm x}^{\rm LDA}[\rho]+E_{\rm c}^{\rm LDA}[\rho])}{\delta\rho(x)}.
\end{equation}
Here
\begin{equation}
	U_{\rm H}[\rho]
	=\frac{1}{2}\int_\mathcal{X}\rho(x)^2dx,
	\qquad\text{and}\qquad
	E_{\rm x}^{\rm LDA}[\rho]
	=-\frac{1}{4}\int_\mathcal{X}\rho(x)^2dx,
\end{equation}
while the correlation energy is approximated by the local density approximation in \cite{magyar2004density}
\begin{equation}
	E_{\rm c}^{\rm LDA}[\rho]
	=\int_\mathcal{X}\frac{a\rho(x)^3+b\rho(x)^2}{\rho(x)^2+d\rho(x)+e}dx,
\end{equation}
where $a=-1/24$, $b=-0.00436143$, $d=0.252758$ and $e=0.0174457$.

We begin by investigating simulations conducted on a single cell. In particular, simulations are performed with varying $\mathbf{x}$ truncation orders $M=8,16,24,32,40,48,56$ and fixed $\mathbf{p}$ truncation orders $N=128$. Using the numerical results in the Schr\"odinger formalism with $M=128$ as the reference, the decay behavior of errors are is illustrated in the left figure of Fig. \ref{fig:1d err-M 1cell}. Additionally, the error decay with respect to the Wigner reference when $M=N=128$ is exhibited in the right figure of Fig. \ref{fig:1d err-M 1cell}. Conversely, by fixing truncation order in the $\mathbf{x}$ direction as $M=128$, Wigner simulations with truncation orders in the $\mathbf{p}$ direction as $N=24,32,40,48,56,64,72$ are accomplished, with error decay depicted in Fig. \ref{fig:1d err-N 1cell}. Similarly, the Schr\"odinger/Wigner reference is considered in the left/right figure of Fig. \ref{fig:1d err-N 1cell}, respectively. Furthermore, the behavior of each component of the total energy is presented separately.
It can be found in Fig. \ref{fig:1d err-M 1cell} that (i) The convergence of density to both the Schr\"odinger reference and Wigner reference, as well as the convergence of energy and energy components, can be obtained in all the simulations. (ii) Spectral convergence with respect to $\mathbf{x}$ truncation order $M$ to both the Schr\"odinger reference and Wigner reference is evident across all the computations. (iii) The error decay of energy components $E_{\rm w,ext}^{N=128}$, $E_{\rm w,kin}^{N=128}$ and $E_{\rm w,c}^{N=128}$ exhibits similar behavior to the density errors, resulting from the fact that these components are determined by the density. (iv) Since the kinetic energy is a functional of all the coefficient functions, its errors show the worst decay behavior compared to others. Particularly, the error of kinetic energy finds no improvement when the truncation order is sufficiently large (e.g., larger than 40), due to the absence of higher coefficient functions. In such cases, discretization errors in the $\mathbf{p}$ direction become dominant.
The numerical observations in Fig. \ref{fig:1d err-N 1cell} can be summarized as follows: (i) Spectral convergence with respect to $\mathbf{p}$ truncation order $N$ to both Schr\"odinger reference and Wigner reference can be obtained in all the simulations. (ii) The errors of density-determined energy components exhibit a similar behavior to the density errors. (iii) Compared to the gentle behavior with sufficient large $N$ using the Schr\"odinger reference, the results with respect to the Wigner reference show a smooth error decay. This is attributed to the use of numerical results with the same $\mathbf{x}$ truncation order $M=128$.

Fig. \ref{fig:1d err 4cell} illustrates the numerical outcomes of simulations conducted on quadruple cells. Specifically, the simulations encompass the $\mathbf{x}$ truncation orders ranging from $M=32$ to $M=224$ with increments of 32, while maintaining a fixed $\mathbf{p}$ truncation order of $N=128$. Additionally, results for $M=256$ with $N=24,32,40,48,56,64,72$ are presented. Throughout these calculations, the Schr\"odinger reference with $M=512$ and the Wigner reference with $M=512,N=128$ are employed to demonstrate convergence behavior. Similarly, the components of total energy are depicted in the figures, facilitating a deeper investigation of the numerical findings.
Similar observations can be obtained in Fig. \ref{fig:1d err 4cell}. (i) The spectral convergence to the reference in both the $\mathbf{x}$ and $\mathbf{p}$ direction can be found in the computations. (ii) The errors of density-determined energy components exhibit a similar behavior to the density errors. (iii) On the contrary, the kinetic energy depends on all coefficient functions, resulting in a gentle behavior of kinetic errors in the top figures when the truncation order is sufficiently large. (iv) Smooth convergence to the reference can be observed in the bottom right figure due the use of numerical Wigner results with the same $\mathbf{x}$ truncation order.

Finally, employing the numerical results in single-cell simulations as a reference, the errors of multiple-cell simulations are presented in Table \ref{tab:1d multi-cell}. In detail, the cell number $n_{\rm cell}=1,2,4,8,16,32,64,2^{20}$ are considered, whose corresponding degree of freedom numbers are shown in the row $n_{\rm DoF}$. The truncation orders are set as $M=32n_{\rm cell}$ and $N=32$. The row $E_{\rm w}$ lists the energy of the corresponding calculation, while $E_{\rm w,avg}$ provides the energy per cell of multiple-cell simulations.
The reference energies $E_{\rm s}^{\rm ref}$, $E_{\rm w}^{\rm ref}$ are the corresponding numerical energies.
And the reference densities $\rho_{\rm s}^{\rm ref}$ and $\rho_{\rm w}^{\rm ref}$ are generated by the periodic extension of the one in the Schr\"odinger formalism with $M=128$ and the one in the Wigner formalism with $M=N=128$, respectively. 
To provide a more comprehensive demonstration of the errors, the average energy errors are presented in the third and fourth rows. Additionally, due to the use of $L_2$ norm, the average density errors, divided by the square root of cell number, are shown in the last two rows.
\begin{table}[H]
	\centering
	\caption{Errors of multiple cell simulations}
	\label{tab:1d multi-cell}
	\footnotesize
	\begin{tabular}{c|cccccccc}\hline
		$n_{\rm cell}$
		&1	&2	&4	&8	&16	&32	&64	&$2^{20}$\\\hline
		$n_{\rm DoF}$
		&1024	&2048	&4096	&8192	&16384	&32768	&65536	&33554432\\\hline
		$E_{\rm w}$
		&1.31	&2.62	&5.24	&10.48	&20.96	&41.92	&83.83	&1373484.49\\
		$|E_{\rm w,avg}-E_{\rm s}^{\rm ref}|$
		&3.91e-03	&3.91e-03	&3.91e-03	&3.91e-03	&3.91e-03	&3.91e-03	&3.91e-03	&3.91e-03\\
		$|E_{\rm w,avg}-E_{\rm w}^{\rm ref}|$
		&3.91e-03	&3.91e-03	&3.91e-03	&3.91e-03	&3.91e-03	&3.91e-03	&3.91e-03	&3.91e-03\\\hline
		$\|\rho_{\rm w}-\rho_{\rm s}^{\rm ref}\|_2$
		&1.25e-04	&1.77e-04	&2.50e-04	&3.54e-04	&5.00e-04	&7.07e-04	&1.00e-03	&2.67e-01\\
		$\|\rho_{\rm w}-\rho_{\rm s}^{\rm ref}\|_2/\sqrt{n_{\rm cell}}$
		&1.25e-04	&1.25e-04	&1.25e-04	&1.25e-04	&1.25e-04	&1.25e-04	&1.25e-04	&1.62e-04\\
		$\|\rho_{\rm w}-\rho_{\rm w}^{\rm ref}\|_2$
		&1.25e-04	&1.77-e04	&2.50e-04	&3.54e-04	&5.01e-04	&7.08e-04	&1.00e-03	&2.67e-01\\
		$\|\rho_{\rm w}-\rho_{\rm w}^{\rm ref}\|_2/\sqrt{n_{\rm cell}}$
		&1.25e-04	&1.25e-04	&1.25e-04	&1.25e-04	&1.25e-04	&1.25e-04	&1.25e-04	&1.62e-04\\\hline
	\end{tabular}
\end{table}
It is revealed in Table \ref{tab:1d multi-cell} that almost all multiple-cell simulations yield identical average energy errors and average density errors, at least to 4 decimal places. Remarkably, the calculation on $2^{20}$ cells produces average energy error and density error of the same order, which also achieves a satisfactory accuracy (less than $1\times10^{-3}$). Conversely, to deliver a similar average error for larger systems, the increasement of the DoF number merely grows linearly w.r.t. the cell number, resulting an almost linear growth of computational complexity for a single iteration. It also is desired mentioned that only a single Wigner function is computed in each simulation, underscoring the descriptive capability of the Wigner formalism and its potential to large scale simulations.

\subsection{3-D Hooke's atom}
\begin{figure}[H]
	\centering
	\includegraphics[width=.48\linewidth]{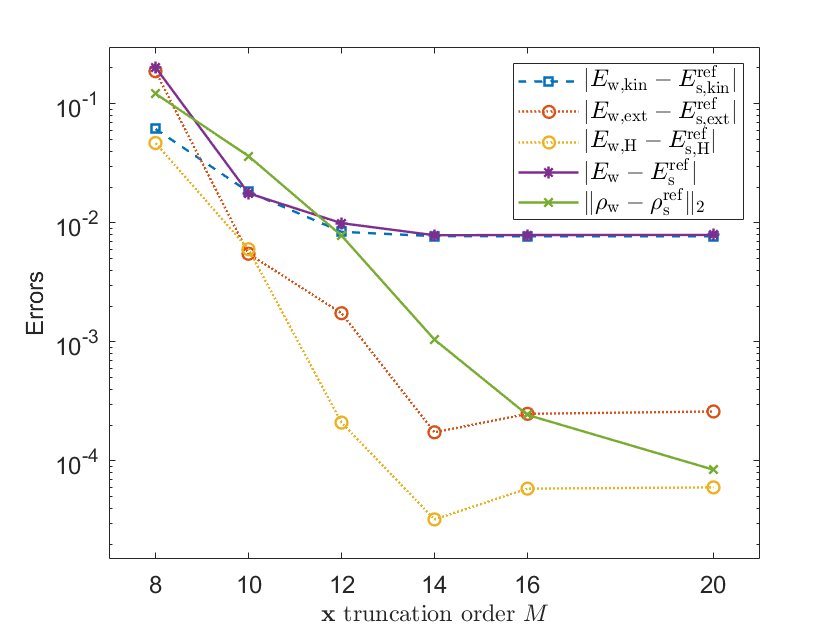}
	\includegraphics[width=.48\linewidth]{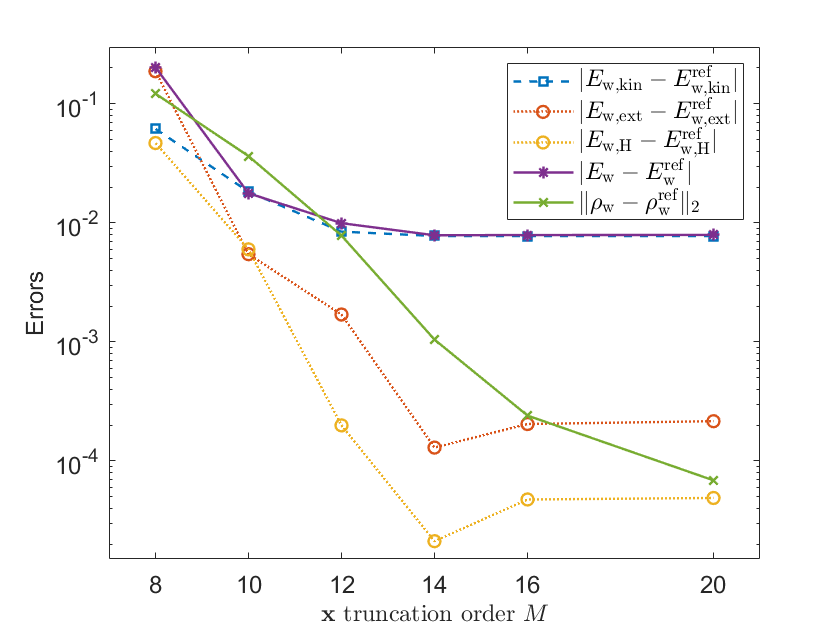}
	\caption{Decay of errors w.r.t. Schr\"odinger (left)/Wigner (left) reference for different $\mathbf{x}$ truncation order in single cell simulations.}
	\label{fig:3d err-M}
\end{figure}

\begin{figure}[H]
	\centering
	\includegraphics[width=.48\linewidth]{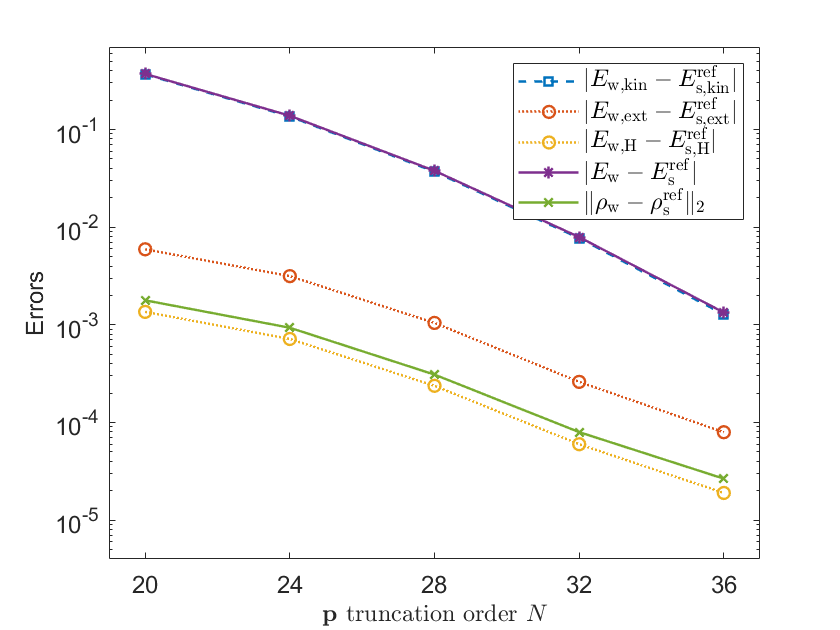}
	\includegraphics[width=.48\linewidth]{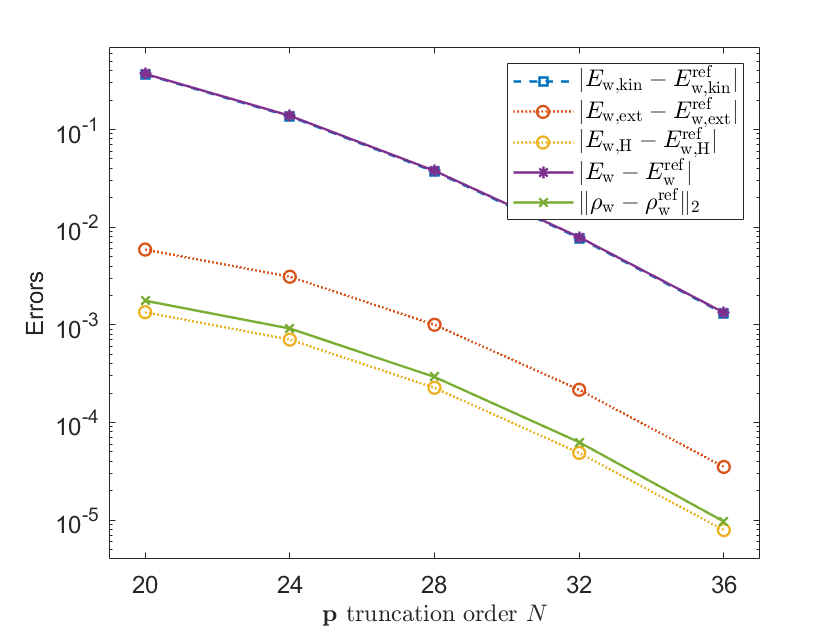}
	\caption{Decay of errors w.r.t. Schr\"odinger (left)/Wigner (left) reference for different $\mathbf{p}$ truncation order in single cell simulations.}
	\label{fig:3d err-N}
\end{figure}

\begin{figure}[H]
	\centering
	\includegraphics[width=.38\linewidth]{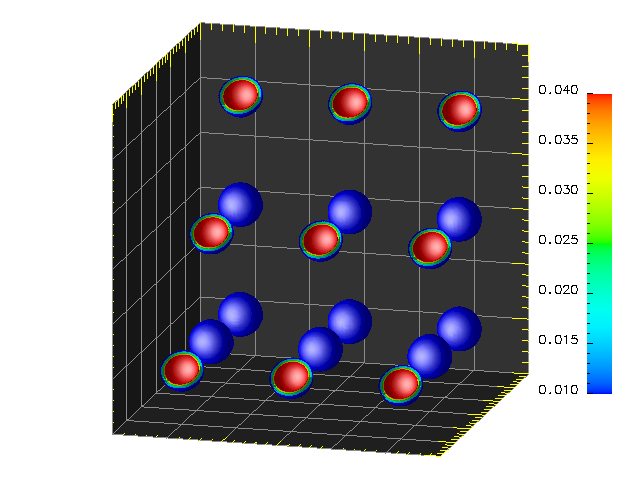}
	\includegraphics[width=.38\linewidth]{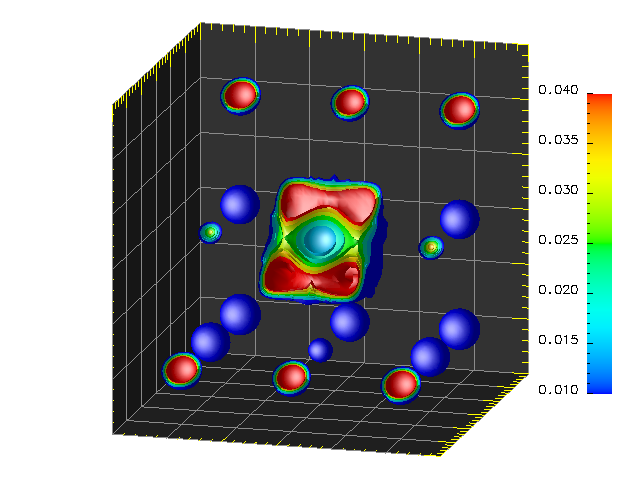}
	\caption{Density distribution of Wigner results for $3\times3\times3$ cells (left) and $3\times3\times3$ with central absence (right).}
	\label{fig:3d den}
\end{figure}

In this sub-section, a three-dimensional system is examined to explore the impact of Coulomb interaction. Firstly, single-cell simulations are accomplished for testifying the spectral accuracy of the Fourier pseudo-spectral method. The results of these computations are collected as Fig. \ref{fig:3d err-M} and Fig. \ref{fig:3d err-N}. Subsequently, both the energy errors and density errors for multiple-cell simulations are exhibited in Table \ref{tab:3d multi-cell}, underscoring the descriptive capability of the Wigner formalism. Moreover, a comparison of density distribution between $3\times3\times3$-cell simulations and those with central absence is demonstrated in Fig. \ref{fig:3d den}, which illustrates the potential applications of our approach for simulating systems with defects.

Particularly, for single-cell system we have
\begin{equation}
	\left(-\frac{1}{2}\nabla^2+V_{\rm ext}(\mathbf{r})+V_{\rm H}[\rho](\mathbf{r})\right)\psi_j(\mathbf{r})
	=\varepsilon_j\psi_j(\mathbf{r}),
\end{equation}
where $\mathbf{r}=(x,y,z)$ stands for three-dimensional coordinates,
\begin{linenomath}\begin{align}
	V_{\rm ext}(\mathbf{r})
	&=\frac{1}{2}|\mathbf{r}|^2,\\
	V_{\rm H}(\mathbf{r})
	&=\int_\mathcal{X}\frac{\rho(\mathbf{r}')}{|\mathbf{r}-\mathbf{r}'|}d\mathbf{r}'.
\end{align}\end{linenomath}
For simplicity, uniform truncation order across all the directions are adopted for the discretization of $\mathbf{x}$ and $\mathbf{p}$ domains, respectively.
We begin with single-cell simulations using $\mathbf{x}$ truncation order $M=8,10,12,14,16,20$ and $\mathbf{p}$ truncation order $N=32$, whose results are demonstrated in Fig. \ref{fig:3d err-M}. Conversely, simulations with $M=32$ and $N=20,24,28,32,36$ are also conducted. The Schr\"odinger results with $M=64$ and the Wigner results with $M=N=48$ are employed as the reference. Similarly, the error behavior of energy components $E_{\rm w,kin}$, $E_{\rm w,ext}$ and $E_{\rm w,H}$ are presently separately for further analysis.
One can observe in Fig. \ref{fig:3d err-M} that (i) Spectral convergence with respect to $M$ to both the Schr\"odinger and Wigner reference can be obtained. (ii) The energy error of density-determined energy components exhibit a similar behavior tp the density errors. (iii) Increasing $M$ shows no improvement of the kinetic energy errors when $M$ is large enough (e.g., $M>32$), resulting from the fact that the discretization error in $\mathbf{p}$ direction dominates in these situations.
It can be found in Fig. \ref{fig:3d err-N} that the density errors, energy errors and the errors of energy components all demonstrate the spectral convergence with respect to $\mathbf{p}$ truncation order $N$.

Next we consider multiple-cell simulations, whose results are concluded as the following table. In each simulation, the $\mathbf{x}$ truncation order is employed as the corresponding multiple of 16, i.e., taking $M_x=16n_x^{\rm multi}$, $M_y=16n_y^{\rm multi}$, $M_z=16n_z^{\rm multi}$ for the cell type $n_x^{\rm multi}\times n_y^{\rm multi}\times n_z^{\rm multi}$, with the $\mathbf{p}$ truncation order $N=32$.
The reference energy and density are obtained from the periodic extension of the Schr\"odinger results with $M=64$ and the Wigner results with $M=N=48$. In Table \ref{tab:3d multi-cell}, the row $E_{\rm w}$ lists the energies of multiple-cell simulations, the row $e_E^{\rm s,avg}$ and the row $e_E^{\rm w,avg}$ list the energy errors per cell with respect to the Schr\"odinger and Wigner reference. Additionally, the row $e_\rho^{\rm s}$ and the row $e_\rho^{\rm w}$ correspond to the density errors with respect to the reference. Similarly, the row $e_\rho^{\rm s,avg}$ and the row $e_\rho^{\rm w,avg}$ exhibit the average density errors in the sense of dividing by the square root of cell number. 
\begin{table}[H]
	\centering
	\caption{Errors of multiple cell simulations}
	\label{tab:3d multi-cell}
	\footnotesize
	\begin{tabular}{c|ccccccccc}\hline
		cell type
		&$1\times1\times1$	
		&$3\times1\times1$	&$1\times3\times1$	&$1\times1\times3$
		&$2\times2\times1$	&$2\times1\times2$	&$1\times2\times2$
		&$2\times2\times2$	&$3\times3\times3$\\\hline
		$n_{\rm DoF}$
		&1.34e08
		&4.03e08		&4.03e08		&4.03e08
		&5.37e08		&5.37e08		&5.37e08
		&1.07e09		&3.62e09\\\hline
		$E_{\rm w}$
		&3.49
		&10.46	&10.46	&10.46	
		&13.95	&13.95	&13.95	
		&27.90	&94.15\\
		$e_E^{\rm s,avg}$
		&7.90e-04	
		&7.90e-04	&7.90e-04	&7.90e-04	
		&7.90e-04	&7.90e-04	&7.90e-04
		&7.90e-04	&7.90e-04\\
		$e_E^{\rm w,avg}$
		&7.90e-04	
		&7.90e-04	&7.90e-04	&7.90e-04
		&7.90e-04	&7.90e-04	&7.90e-04
		&7.90e-04	&7.90e-04\\\hline
		$e_\rho^{\rm s}$
		&2.43e-04	
		&4.21e-04	&4.21e-04	&4.21e-04
		&4.85-e04	&4.85e-04	&4.85e-04
		&6.86e-04	&1.30e-03\\
		$e_\rho^{\rm s,avg}$
		&2.43e-04
		&2.43e-04	&2.43e-04	&2.43e-04
		&2.42e-04	&2.42e-04	&2.42e-04
		&2.42e-04	&2.50e-04\\
		$e_\rho^{\rm w}$
		&2.40e-04
		&4.16e-04	&4.16e-04	&4.16e-04
		&4.80e-04	&4.80e-04	&4.80e-04
		&6.78e-04	&1.28e-03\\
		$e_\rho^{\rm w,avg}$
		&2.40e-04
		&2.40e-04	&2.40e-04	&2.40e-04
		&2.40e-04	&2.40e-04	&2.40e-04
		&2.40e-04	&2.46e-04\\
		\hline
	\end{tabular}
\end{table}
It can be found in Table \ref{tab:3d multi-cell} that multiple-cell simulations yield consistent average energy errors and average density errors in all the cell type.  Moreover, all the average errors are less than $1.0\times 10^{-3}$, deliver a desired accuracy.
Similarly, a almost linear increasement of the computational complexity for a single iteration, resulting from the linear growth of the DoF number w.r.t. the system scale, delivers a comparable average error as the single cell case.
It is worth mentioning that the Wigner formalism enables the description of the entire system by a single Wigner function regardless of the system size. Moreover, in our method, the computational cost of single evolution increases almost linearly with the system size. This underscore the potential of our approach for large-scale simulations.

Finally, the visualization results of $3\times3\times3$ cell simulations are exhibited in Fig. \ref{fig:3d den}. The density distribution of the system without central cell arrangement is also included. In particular, the $\mathbf{x}$ truncation order is employed as $M_x=M_y=M_z=48$, and the $\mathbf{p}$ truncation order is $N=16$ in all directions. In the simulation corresponding to the left figure of Fig. \ref{fig:3d den}, each sub-cell subjects to the external potential $V_{i_{\rm cell}}(\mathbf{r})=|\mathbf{r}-\mathbf{r}_{\rm center}^{i_{\rm cell}}|^2/2$, where $i_{\rm cell}$ is the cell index, and $\mathbf{r}_{\rm center}^{i_{\rm cell}}$ is the center position of this sub-cell. On the contrary, the simulation shown in right figure features the absence of the external potential in center sub-cell, resulting in the electron number 54 compared to the one 56 in the left figure. To better illustrate the density distribution, we consider the isosurfaces for the density $\rho=0.01+0.005j$ with $0\le j\le6$, clipped by the plane $y=z$.
It can be observed in Fig. \ref{fig:3d den} that (i) in the left figure, the density within each sub-cell exhibits the same distribution, which is symmetric with respect to the sub-cell center. (ii) Similarly, the density distribution in sub-cells of the right figure also displays symmetric behavior. (iii) Specially, due to the absence of external potential in the right figure, only few density is present around the cell center of the entire system. Additionally, the sub-cells near to the center sub-cell also exhibit smaller density distributions.
It is worth mentioning that the only difference between the numerical simulations for these two figures is the external potential and the electron number. Therefore, Wigner computations are expected to offer a more straightforward description of complex systems in numerical calculations. This arises from the feature that Kohn-Sham orbitals are expressed as an ensemble in the corresponding Wigner function within the Wigner formalism.

\section{Conclusion}\label{sec:conclusion}
In this paper, a gradient flow model is derived for Wigner ground state calculation of many-body system in the context of density functional theory. In particular, a gradient flow model for one-body systems is firstly derived and then extended to many-body systems within the framework of density functional theory. To enhance computational efficiency, an operator splitting scheme and the Fourier pseudo-spectral method are introduced for numerical simulations, which delivers a parallelizable algorithm with $O(n_{\rm DoF}\log n_{\rm DoF})$ computational cost for a single evolution step. Two toy models, based on the periodic extensions of two-body systems, are presented to demonstrate the efficacy of our approach, encompassing a one-dimensional delta-interacting example with a local density approximation and a three-dimensional system with Coulomb interaction. Spectral accuracy can be successfully observed in computations, while multiple-cell simulations recover the desired errors compared to the single-cell simulations. Moreover, the use of the same setup in $\mathbf{p}$ space for multi-cell calculations results in an almost linear increasement in computational complexity for a single iteration, demonstrating the potential of the proposed method for simulating large-scale systems and systems with defects.

As for future work, there are two primary avenues of exploration. On the one hand, we aim to extend our simulations to encompass more general systems, such as complex molecular systems. On the other hand, we aim to address the current limitations in the size of the discretized system required for accurately representing three-dimensional systems. To enhance the efficiency of our method, we will explore numerical strategies, such as selectively ignoring certain coefficient functions to reduce computational resource requirements per iteration.
Alternatively, we may explore the Grad moment method, which possesses a straightforward expressions of both density and energy.

\bibliographystyle{plain}
\bibliography{ref}

\end{document}